\documentclass
[11pt]{article}
\begin{document}
\begin{center}
{\Large\bf Ernst Ising - physicist and teacher}\\

\vspace{0.6cm}
S. KOBE   \\
\vspace{0.1cm}
{\it Technische Universit\"at Dresden, 
Institut f\"ur Theoretische Physik, \\ D-01062 Dresden, Germany }
\end{center}
\vspace{0.4cm}
{\small The Ising model is one of the standard models in statistical physics.
Since 1969 more than 12000 publications appeared using this model.
In 1996 Ernst Ising celebrated his 96th birthday. Some biographical
notes and milestones of the development of the Ising model are given.} \\

\vspace{0.5cm}
Ernst Ising was born on May 10, 1900 in K\"oln (Cologne, Germany), the son
of the merchant Gustav Ising and his wife Thekla, nee L\"owe.
The family moved to Bochum (Westfall). Ernst Ising completed the
Gymnasium in Bochum in 1918. Shortly after that he had to do a brief military
training. In 1919 he started studying mathematics and physics at the
University of G\"ottingen and continued his studies in Bonn and Hamburg.
In Hamburg Wilhelm Lenz (1888 - 1957) suggested to Ising to turn to
theoretical physics. Under his guidance he began
investigating a model of ferromagnetism which was
introduced by his supervisor in 1920 \cite{lenz}.
In his dissertation (1924) Ising studied the special case of a
linear chain of magnetic moments, which are only able to take two positions
''up'' and ''down'' and which are coupled by interactions between
nearest neighbours. He showed that spontaneous magnetization
cannot be explained using this model in its one-dimensional version
\cite{ising}. \\

\smallskip

After receiving his doctor's degree Ising went to Berlin and
worked from 1925 to 1926 in the patent office of the Allgemeine
Elektrizit\"atsgesell- schaft (AEG). He was not satisfied
with this job and decided to become a teacher. For a year
he worked at the famous boarding school in Salem in South-Baden
(near the lake Constance). In 1928 he returned to Berlin University
to study philosophy and pedagogy. In 1930 he passed the state
examinations before the school board in higher education. In the same
\linebreak --------------------------\\
%\vspace{0.05cm}\\
{\footnotesize published in:
Actas: Noveno Taller Sur de Fisica del Solido, Misi\'on Boroa, Chile, 
26-29 April 1995 (Universidad de la Frontera, Temuco, Chile, 1995) p.1.\\
The author is indebted to K. Puech for the critical reading of the 
manuscript.}
\newpage
\noindent year he married Johanna Ehmer, a doctor of economy.
The couple went to Strausberg near Berlin, where Ernst Ising
got a teaching position at a high school as ''Studienassessor"
(in Germany a holder of a higher civil service post,
who has passed the necessary examinations but
has not yet completed his probationary period). Later he was tranferred to
Crossen on the river Oder (now: Krosno in Poland) to fill in for an ill
colleague. \\

\smallskip
Just after Hitler came to power in January 1933, Jewish citizens were
dismissed from their jobs as civil servants. Consequently, Ernst Ising lost 
his position on March 31st, 1933. For one year he was unemployed, except for
a short stay in Paris at a school for emigrant children.
In 1934 he found a new job as a teacher at a boarding school
for Jewish children in Caputh near Potsdam. This school was founded
by the progressive social educationalist Gertrud Feiertag (1890 - 1942?).
The number of students increased in the following years, because Jewish
children had been dismissed from public schools. In 1937 Ising became the
headmaster of this school. In this position
on November 10, 1938 he experienced the devastation of
his school building by incited inhabitants and
children of the village as part of the great pogrom against
the Jewish population in Germany.   \\

\smallskip
In the beginning of the year 1939 Ernst and Johanna Ising traveled
to Luxembourg with the plan to emigrate to the United States of
America. But at that time there was a quota of immigration to the U.S.A. and
so it was impossible for them to fulfil their plan. On Ising's fortieth
birthday the German army occupied Luxembourg. The U.S. consulates
were closed just before the waiting time for emigration was over.
Ising and his family survived the war. From April 1943 until the liberation
in 1944, he was forced to work for the German army. \\

\smallskip
Two years after the end of the war Ising left Europe on board a
freighter to the U.S.A. He got a job as a teacher at the State Teacher's 
College in Minot (N.D.). From 1948 till 1976 he was Professor of Physics
at the Bradley University Peoria (Il.), earned the Honorary Doctor's
degree of this university in 1968 and was named ''Outstanding
Teacher of the Year 1971". Since 1953 Ernest Ising is U.S. citizen.
After his retirement he traveled a lot and visited many countries all
over the world. \\

\newpage

Now Ising lives in Peoria. Recently he celebrated many jubilees,
among them were the 70th anniversaries of his doctor's degree, the
appearance of his publication \cite{ising}, his 96th birthday, and his
iron wedding anniversary. \\

\smallskip
Ising became aware of the first citation of his paper by Heisenberg
\cite{heisen}, who introduced the quantum mechanical exchange
interaction to describe ferromagnetism.
Later Ising was completely shut off from scientific life
and communication for a long time. It was not until 1949
that Ising found out from the scientific literature that his paper
had become widely known. The main drive for calling the model ''Ising
model'' seems to stem from R. Peierls' publication
''On Ising's Model of Ferromagnetism'' \cite{peierls}. He argued that
spontaneous magnetization must exist, but his proof was not quite rigorous.
This was shown later by N.G. van Kampen, M.E. Fisher, S. Sherman
and others, cf. \cite{brush}, \cite{griffiths}. Further history
of the model is characterized by the search for the phase transition between
the ferromagnetic and the paramagnetic state, cf. \cite{brush}.
The actual breakthrough came from the findings of various authors
\cite{krwa} - \cite{kubo} which say that a matrix representation of the
problem can be introduced in such a way that the partition function can be
related to the largest eigenvalue of this matrix. Kramers and Wannier
\cite{krwa} have calculated the numerical value for the Curie temperature
of the two-dimensional version of the Ising model, whereas the exact
and complete solution was first given by Onsager \cite{ons}. \\

\smallskip
Today the Ising model is a widely used standard model of statistical
physics. Every year 700 - 900 papers which apply this
model are published; problems regarding neural networks, protein
folding, biological membranes, social imitation and frustration are
among them. \\

\smallskip
In this contribution some biographical notes on the life and work
of E. Ising are given and the early history of
the Ising model is mentioned. Further biographical sources can be found in
\cite{brush}, \cite{fry} - \cite{jising}. \\

\newpage
\section*{Acknowledgements}
I am deeply indebted to J. Ising for letting me have and use her unpublished
memoirs \cite{jising}. I thank M. Suzuki for drawing my attention to Kubo's
paper \cite{kubo}. \\

\end{document}